\documentclass[12pt]{iopart}
\usepackage[dvips]{graphicx}
\usepackage{color}
\usepackage{iopams}

\begin{document}

\title{Quantum Monte Carlo Study of High Pressure Solid Molecular Hydrogen} 

\author{}

\author{Sam Azadi$^1$, W. M. C. Foulkes$^1$, Thomas D. K\"{u}hne$^2$}
\address{$^1$ Thomas Young Centre and Department of Physics, Imperial College London,  
London SW7 2AZ, United Kingdom}
\address{$^2$ Institute of Physical Chemistry and                                                  
Centre for Computational Sciences, Johannes Gutenberg University
Mainz, Staudinger Weg 9, 55128 Mainz, Germany}
\ead{s.azadi@imperial.ac.uk}


\begin{abstract}
  We use the diffusion quantum Monte Carlo (DMC) method to calculate the
  ground state phase diagram of solid molecular hydrogen and examine the
  stability of the most important insulating phases relative to metallic
  crystalline molecular hydrogen.  We develop a new method to account
  for finite-size errors by combining the use of twist-averaged boundary
  conditions with corrections obtained using the Kwee-Zhang-Krakauer
  (KZK) functional in density functional theory. To study band-gap
  closure and find the metallization pressure, we perform accurate
  quasi-particle many-body calculations using the $GW$ method.  In the
  static approximation, our DMC simulations indicate a transition from
  the insulating Cmca-12 structure to the metallic Cmca structure at
  around 375 GPa. The $GW$ band gap of Cmca-12 closes at roughly the
  same pressure. In the dynamic DMC phase diagram, which includes the
  effects of zero-point energy, the Cmca-12 structure remains stable up
  to 430 GPa, well above the pressure at which the $GW$ band gap
  closes. Our results predict that the semimetallic state observed
  experimentally at around 360 GPa [Phys. Rev. Lett. {\bf 108}, 146402
  (2012)] may correspond to the Cmca-12 structure near the pressure at
  which the band gap closes.  The dynamic DMC phase diagram indicates
  that the hexagonal close packed $P6_3/m$ structure, which has the
  largest band gap of the insulating structures considered, is stable up
  to 220 GPa. This is consistent with recent X-ray data taken at
  pressures up to 183 GPa [Phys. Rev. B {\bf 82}, 060101(R) (2010)],
  which also reported a hexagonal close packed arrangement of hydrogen
  molecules.

\end{abstract}

\maketitle

\section{Introduction}

Solid molecular hydrogen is the simplest and most fundamental quantum
molecular crystal. At high pressure it is predicted to become not only a
metallic atomic solid \cite{wigner} but also a high-$T_c$ superconductor
\cite{Ashcroft1} and perhaps a superfluid \cite {Ashcroft2}. Although
there are experimental results for the structure and phase diagram of
high-pressure hydrogen at room temperature, \cite {Loubeyre1, Loubeyre2,
  Goncharov-PNAS, Eremets1, Eremets2} the crystal structures appearing
in the complex low-temperature high-pressure phase diagram are not yet
completely determined.  Spectroscopic investigations using infra-red
(IR) and Raman methods suggest the existence of three
phases \cite{Hemley}. Phase I, which is stable up to 110 GPa, is a
molecular solid composed of quantum rotors arranged in a
hexagonal-close-packed (hcp) structure.  Changes in the low-frequency
regions of the Raman and IR spectra imply the existence of a second
phase above 110 GPa; this is known as phase II, or the broken symmetry
phase.  The continuous behavior of the Raman vibron frequency across the
I to II transition suggests that phase II may consist of an arrangement
of molecules on an hcp lattice, as in the P$6_3$/m
structure. Calculations based on density functional theory (DFT) predict
orthorhombic Cmc$2_1$, Pca$2_1$, and P$2_1$/c
structures \cite{Kitamura, Kohanoff1, Johnson}. The molecular centres in
these structures lie close to those of an hcp lattice, but are
incompatible with neutron data for solid deuterium \cite{Loubeyre3}. The
appearance of phase III at 150 GPa is accompanied by a large
discontinuity in the Raman spectrum and a strong rise in the spectral
weight of the IR molecular vibrons.  Although the structures of phase II
and phase III are not known for certain, recent X-ray results
\cite{Akahama} at pressures up to 183 GPa suggest that the hydrogen
molecules in both phases lie close to the points of an hcp
lattice. Recent infrared and optical absorption results \cite{Hemley12}
indicate that the molecular structure of phase III persists up to 360
GPa, with no evidence of the onset of a metallic state.

Computational methods used to investigate the high-pressure
low-temperature phase diagram of solid molecular hydrogen include
density functional theory \cite{Johnson, Martin, Kohanoff1, Kohanoff2,
  Chacham, Nagao, JETP} and quantum Monte Carlo (QMC) \cite{Natoli1,
  Natoli2}. Approaches used to explore the phase space of possible
structures include the evolutionary method of Oganov and Glass
\cite{Oganov} and the static ab-initio structure searching method of
Pickard and Needs, \cite{Pickard} which has also been employed to
determine the ground-state structure of \emph{atomic} metallic
hydrogen \cite{McMahon}. Recently, a series of four papers \cite{JCP1,
  JCP234} presented a fresh look at the high-pressure hydrogen problem.
By studying the equalization of the intra- and inter-molecular H-H bond
lengths in dense hydrogen, these papers investigated the four most stable
structures found by Pickard and Needs \cite{Pickard} using DFT in the
static approximation (i.e., ignoring the zero-point vibrational energy
of the H atoms).

However, all of the recent calculations of the metallization pressure
and phase diagram are based on the single-particle mean-field-like DFT
method. This has various drawbacks, such as the band-gap problem and the
sensitivity of the results to the choice of exchange-correlation (XC)
functional, which affects both the relative stabilities of different
phases and the predicted metallization pressure \cite{Azadi}.
Furthermore, attempts to explore the phase space of possible structures
have been based primarily on DFT in the static approximation, which
ignores the proton zero-point energy (ZPE).  Since the ZPE is normally
larger than the energy differences between various candidate structures,
structure-searching methods including an approximate ZPE should be
considered. Most of the existing DFT calculations of solid hydrogen have
used the local density approximation (LDA) or the generalized gradient
approximation (GGA) to the unknown exchange-correlation
functional. Neither of these functionals adequately describes van der
Waals (vdW) interactions.


In this paper, we use QMC simulations to determine the ground-state
phase diagram of solid molecular hydrogen in the pressure range $100 < P
< 400$ GPa.  We study, systematically and accurately, the two main
factors affecting the quality of the calculated phase diagram:
finite-size errors and proton zero-point energy. To investigate the
mechanism of metallization and determine the pressure at which the band
gap closes, we also carry out many-body perturbation theory calculations
within the $GW$ approximation.  We focus on four specific structures,
with space groups P$6_3$/m, C2/c, Cmca-12, and Cmca. According to
ground-state static DFT calculations using the Perdew-Burke-Ernzerhof
(PBE) version of the generalized gradient approximation \cite{pbe}
(GGA), these are stable in the pressure ranges $<$105, 105--270,
270--385, and 385--490 GPa, respectively \cite{Pickard}.

\section{Computational Details}

The fixed-node diffusion quantum Monte Carlo (DMC) algorithm \cite{rmp}
yields the energy expectation value of the variationally optimal wave
function with the same nodal surface (surface of zeros) as a chosen
many-electron trial function. The method has a long history and has been
thoroughly benchmarked \cite{rmp}. For small molecules, where almost
exact results can be obtained using quantum chemical methods,
ground-state total energies calculated using fixed-node DMC are
consistently more accurate than energies calculated DFT, typically by a
factor of 10 or more. Unlike DFT, fixed-node DMC also provides a good
description of the dependence of the total energy on coordination number
(very important when comparing the energies of different crystal
structures) and a fairly good description of the potential surface
during bond-breaking processes. For large systems such as the ones
studied in this paper, fixed-node DMC is the most accurate known
ground-state total energy method and the only possible benchmarks are
against experiment. Again, however, DMC consistently outperforms
DFT. For example, the measured cohesive energy of crystalline Si is 4.62
eV/atom; fixed-node DMC simulations yields 4.63 eV/atom \cite{leung1999};
and DFT in the local spin-density approximation yields 5.28 eV/atom. In
this unusually favorable case, the DMC result is 65 times more accurate
than the DFT result. The ability of DMC to describe changes in
coordination is illustrated by the example of the Si self-interstitial
defect \cite{leung1999}. The DMC value of the defect formation energy,
which is in good agreement with experiment, is about 1 eV larger than
the PW91 GGA value and 1.5 eV larger than the LDA values. The main
drawback of DMC, which explains why it is not more widely used, is the
computational cost: a DMC simulation typically costs about $10^4$ times
more than a DFT simulation of the same system.

DMC can also be used to obtain limited information about excited
states. However, other approaches such as the $GW$ method used in this
work are generally better for excitation spectra. The results of $GW$
calculations of the band gaps of $sp$-bonded materials generally match
experiment to within a few tenths of an eV. 

Our fixed-node DMC simulations used the CASINO QMC code
\cite{casino} and a trial function of Slater-Jastrow (SJ) form,

\begin{equation*}
\Psi_{SJ}({\bf R})=\exp[J({\bf R})]\det[\psi_{n}({\bf r}_i^{\uparrow})]\det[\psi_{n}({\bf r}_j^{\downarrow})] ,
\end{equation*}
 
where ${\bf R}$ is a $3N$-dimensional vector that defines the positions
of all $N$ electrons, ${\bf r}_i^{\uparrow}$ is the position of the i'th
spin-up electron, ${\bf r}_j^{\downarrow}$ is the position of the j'th
spin-down electron, $\exp[J({\bf R})]$ is the Jastrow factor, and
$\det[\psi_{n}({\bf r}_i^{\uparrow})]$ and
$\det[\psi_{n}({\bf r}_j^{\downarrow})]$ are Slater determinants
of spin-up and spin-down one-electron orbitals.  These orbitals were
obtained from DFT calculations using the plane-wave-based Quantum
Espresso code \cite{QS}. A norm-conserving pseudopotential constructed
within DFT using the PBE exchange-correlation functional was
employed. \cite{Marzari} Previous work has shown that the
pseudopotential approximation has a very small impact on the results of
calculations of high-pressure solid hydrogen \cite{McMahon}. We chose a
very large basis-set cut-off of 300 Ry to guarantee converge to the
complete basis set limit \cite{sam}. The plane-wave orbitals were
transformed into a blip polynomial basis \cite{blip}. Our Jastrow factor
consists of polynomial one-body electron-nucleus (en) and two-body
electron-electron (ee) terms, the parameters of which were optimized by
variance minimization at the variational Monte Carlo (VMC)
level \cite{varmin1,varmin2}. 

Geometry and cell optimizations were carried out using DFT with a dense
$16\times16\times16$ {\bf k}-point mesh.  The BFGS quasi-Newton
algorithm, as implemented in the Quantum Espresso package, was used for
both cell and geometry optimization, with convergence thresholds on the
total energy and forces of 0.01 mRy and 0.1 mRy/Bohr, respectively, to
guarantee convergence of the total energy to less than 1 meV/proton and
the pressure to better than 0.1 GPa.  An investigation of the
proton-proton distance of accurately relaxed structures in the pressure
range of 100-450 GPa reveals two main features.  First, as the pressure
increases, the intra-molecular H$_2$ bond length (nearest-neighbor
distance) increases slightly, rising from 0.735\,\AA\  to
0.778\,\AA. This could be a consequence of a weakening of the H$_2$
bonds as the coordination of the H atoms increases. Second, unlike the
intra-molecular H$_2$ bond length, the inter-molecular distance
decreases smoothly as the pressure rises. Details of the behavior of the
inter-atomic distance with pressure and of the experimental evidence are
presented in Ref \cite{JCP1}.

\section{Results and discussion}

\subsection{Effect of single-particle Slater determinant on QMC energy}

Our recent DFT work \cite{JETP} showed that hybrid exchange-correlation
functionals including a small fraction of exact exchange favor
insulating phases relative to metallic phases. We therefore investigate
how the choice of density functional used to calculate the one-electron
orbitals appearing in the Slater determinants (and thus to define the
fixed nodal surface) affects our QMC results.  For this purpose, we
compared the VMC and DMC energies of the insulating P$6_3$/m phase
(which has the largest band gap of the four structures considered) using
orbitals calculated using the semi-local PBE GGA functional \cite{pbe}
and the PBE0 hybrid functional including 25\% exact
exchange \cite{Adamo}. The VMC and DMC calculations were performed at
the $\Gamma$ point of the Brillouin zone of a simulation cell containing
128 hydrogen atoms and $2\times2\times2$ replicas of the P$6_3$/m unit
cell.  The VMC and DMC energies at pressures 100 GPa, 200 GPa and
300 GPa, are given in \Tref{PBE0}.  

\begin{table}
\caption{\label{PBE0} VMC and DMC energies of the P$6_3$/m phase calculated 
    using PBE and PBE0 single-particle orbitals. Energies and 
    pressures are in eV per proton and GPa, respectively.}
\begin{indented}
\item[]\begin{tabular}{@{}lllll}
\br
 Pressure & PBE-VMC & PBE0-VMC & PBE-DMC & PBE0-DMC\\
\mr
 100  & -15.1335(4) & -15.1397(4) & -15.2327(3) & -15.2338(3) \\
 200  & -14.5898(4) & -14.5979(4) & -14.6856(3) & -14.6867(3) \\
 300  & -14.0473(4) & -14.0524(4) & -14.2184(3) & -14.2193(3) \\
\br
\end{tabular}
\end{indented}
\end{table}

The difference between the PBE
and PBE0 energies is about 6 meV/atom at the VMC level and 1 meV/atom at
the DMC level.  Similar calculations for the Cmca-12 structure at two
different pressures gave similar results. This demonstrates that, in the
case of solid molecular hydrogen, DMC energies calculated using PBE and
PBE0 orbitals differ negligibly, even though the change of functional
has a large effect on the DFT phase diagram \cite{JETP}. The very weak
dependence of the DMC energy on the choice of single-particle orbitals
suggests, furthermore, that the fixed-node errors are small. This
contrasts with the case of 3d transition metal compounds \cite{mitas},
where changing the DFT exchange-correlation functional does affect the
DMC results.

\subsection{Finite-size correction to the DMC energy }

When simulating infinite periodic systems or finite systems subject to
periodic boundary conditions, it is not possible to use the familiar
$1/r$ form of the Coulomb interaction because the sums over images of
the simulation cell do not converge absolutely. The standard solution to
this problem is to replace the Coulomb interaction by the Ewald
interaction. The three-dimensional Ewald interaction is the periodic
solution of Poisson's equation for a periodic array of point charges
embedded in a uniform neutralizing background and is therefore
appropriate for an electrically unpolarized, neutral
system \cite{Ewald}. All DMC results reported in this paper were
calculated using the Ewald \cite{Neil} electron-electron interaction.

In DFT, the limit of infinite system size can be approached by improving
the ${\bf k}$-point sampling, but QMC calculations have to be performed
using increasingly large simulation cells.  The resulting finite-size
(FS) effects are a significant source of error.  The following QMC
results used C2/c, Cmca-12, P$6_3$/m and Cmca simulation cells
containing 96, 96, 128, and 144 atoms, respectively, equivalent to
$2\times2\times2$, $2\times2\times2$, $2\times2\times2$ and
$3\times2\times3$ repetitions of the corresponding unit cells.  QMC FS
errors are conventionally separated into one-body and two-body
contributions.  One-body (independent-particle) errors arise from the
non-interacting kinetic, potential and Hartree energies; two-body errors
arise from the effects of exchange and correlation on the Coulomb and
kinetic energies.

This work accounts for FS errors by correctly and accurately combining
the use of twist-averaged boundary conditions (TABC) \cite{twistav} with
DFT-based corrections obtained using the size-dependent
Kwee-Zhang-Krakauer (KZK) functional \cite{kzk}. By construction, KZK
corrections remove the finite-size errors in DMC calculations of cubic
homogeneous systems exactly; our systems are neither cubic nor
homogeneous, but we show below that the KZK approach still works
remarkably well. Our DMC-TABC energies were averages over 12 random
twists, corresponding to 12 randomly translated Monkhorst-Pack ${\bf
  k}$-point meshes. The KZK correction is given by $FS^{KZK} =
E^{LDA}(\infty) - E^{KZK}(L)$, where $E^{LDA}(\infty)$ is the DFT energy
computed within the local density approximation (LDA) using a fully
converged {\bf k}-point mesh (which in this work means
$24\times24\times24$), and $E^{KZK}(L)$ is obtained from DFT-KZK
calculations employing the same simulation cell and twists as our
DMC-TABC results.  In the case of the metallic Cmca structure, we also
carried out DMC-TABC calculations using 24 twists; the change in the
total energy was about 1.4 meV/proton at $P=400$ GPa, implying that DMC
energies obtained by combining TABC and KZK FS-corrections converge very
rapidly as the number of twists is increased.

\begin{table}
\caption{\label{TAVBC} DMC energies of the Cmca phase at P = 300 GPa for different number $N$ of particles
  in simulation cell, N. FS corrections obtained using combination of TABC 
   and DFT-based KZK functional has been applied to the DMC results. Energies are in eV per proton.}
\begin{indented}
\item[]\begin{tabular}{@{}ll}
\br
 N & DMC-TABC \\
\mr
 64  & -14.1829(6) \\
 96  & -14.1931(3) \\
 128 & -14.1936(3) \\
 144 & -14.1927(4) \\
\br
\end{tabular}
\end{indented}
\end{table}

To check the accuracy and convergence of the DMC energy as a function of
the number of protons $N$ in the simulation cell, we performed
calculations for different $N$.  \Tref{TAVBC} shows the results for
the metallic Cmca phase at $P=300$ GPa. The convergence of the TABC-DMC
energy with increasing $N$ is excellent and there is no difficulty in
reaching the accuracy of 1\,meV per proton necessary for high-pressure
hydrogen phase diagram calculations.  We emphasize the importance of our
approach to the treatment of finite-size errors as an application of
TABC to real systems \cite{SAMFS}.

\subsection{Metallization pressure}

To find the metallization pressure, quasi-particle band gaps were
calculated within the $GW$ approximation using the Yambo code \cite{yambo}. 
Non-self-consistent $GW$ calculations were performed
starting from DFT wave functions, using a plasmon pole approximation for
the dielectric function \cite{Hedin, Rubio}. Following convergence
checks, the numbers of unoccupied bands and ${\bf k}$-points were set to
200 and 512, respectively. We also checked convergence with respect to
the number of plane waves, which was set to 1800 for the P$6_3$/m
structure and 2700 for the Cmca-12 and C2/c structures. It is important
to point out that our $GW$ and DMC results are independent of one
another. DMC is used to calculate the enthalpy as a function of
pressure, whist $GW$ is used to calculate the quasi-particle band gap as
a function of pressure. Both methods use the same relaxed DFT crystal
structures, but they are otherwise unrelated.

\begin{figure}
\centering
\includegraphics[width=0.6\columnwidth]{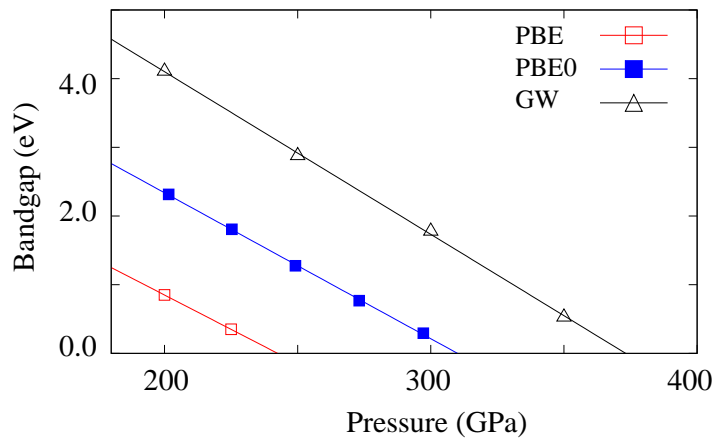}
\caption{
\label{Cmca12}
(Color online) Band gap of the Cmca-12 crystal structure as a function
of pressure as calculated using the PBE, PBE0, and $GW$ approximations.}
\end{figure}

In the following, we present the metallization pressure of the Cmca-12
phase as obtained using the PBE, PBE0, and $GW$ approximations.  We use
the simplest estimate of the metallization pressure, which is the
pressure at which the energy band gap closes. To estimate the
metallization pressure from the calculated band gaps at a few different
pressures, we linearly extrapolate the band gap as a function of
pressure.  Figure \ref{Cmca12} illustrates band-gap closure in the
Cmca-12 phase.  The PBE, PBE0, and $GW$ results show that the band gap
of Cmca-12 vanishes at pressures of 245, 309, and 373 GPa, respectively.
The PBE metallization pressure is in agreement with the value estimated
by Pickard and Needs \cite{Pickard}. From band-structure calculations we
see that the Cmca-12 phase has an indirect band gap; the linear decrease
of the band gap with pressure is not unexpected in indirect gap
structures and is also reported experimentally \cite{Hemley}.

As usual, band gaps calculated using the semi-local PBE functional are
much too small. A significant contributor to this band-gap
underestimation is the inability of \emph{any} analytic functional of
the density to represent the discontinuous change in
exchange-correlation potential that occurs as a single electron is added
to a system with a filled band \cite{sham}. By including a fraction of
the exact non-local orbital-dependent exchange energy, the PBE0 hybrid
functional circumvents this problem to some extent. However, even the
PBE0 band gap is considerably smaller than the $GW$ band gap, which is
expected to be more accurate.

Our $GW$ results show that the band gaps of the P$6_3$/m, C2/c, and
Cmca-12 structures vanish at the static pressures of 446, 398, and 373
GPa, respectively.  Semilocal DFT, hybrid-DFT \cite{JETP}, and $GW$
calculations all show that P$6_3$/m has the largest band gap and,
consequently, the largest metallization pressure of the three insulating
structures considered. The $GW$ indirect band gap and its pressure
dependence agree very well with recently reported experimental
results \cite{Goncharov-PRB}. 

\subsection{Static DMC phase diagram}
\label{StaticDMC}

\begin{figure}
\centering
\includegraphics[width=0.50\columnwidth, angle=-90]{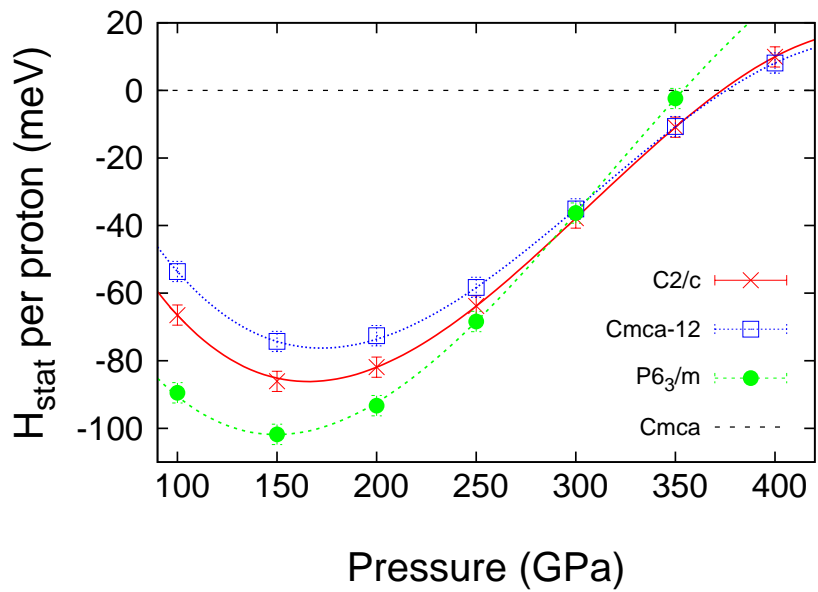}
\caption{\label{H-stat-P-stat}(color online). Enthalpy as a function of
  pressure calculated using the DMC-TABC method in the static
  approximation.}
\end{figure}

In the following, we present enthalpy-pressure diagrams obtained from
DMC simulations at the static level of theory, which ignores proton
ZPE. Figure ~\ref{H-stat-P-stat} illustrates $H_{stat}$ as a
function of pressure. Within the static approximation, the P$6_3$/m
structure is the most stable up to pressures of approximately 250 GPa,
but is less stable than both C2/c and Cmca-12 by 350 GPa. The error bars
make it difficult to establish exactly where the transition from
P$6_3$/m to C2/c or Cmca-12 takes place or to say which of C2/c or
Cmca-12 is favored. The transition from C2/c or Cmca-12 to the metallic
Cmca phase happens around 375 GPa.

Thus, within the static approximation, DMC predicts that the hexagonal
close packed P$6_3$/m structure is more stable than the other structures
investigated over the entire pressure range at which phases I and II
exist experimentally, and that it remains more stable well above the
measured transition to phase III. As already discussed above, recent
X-ray results \cite{Akahama} at pressures up to 183 GPa suggest that hcp
symmetry is a strong candidate not only for phase II but also for phase
III. 

We claim that our static DMC phase diagram is much more accurate than
the static DFT phase diagram. One reason is the importance of vdW
interactions between hydrogen molecules, which significantly influence
the phase diagram, especially at low densities. QMC calculations have
recently shown that the vdW interaction has a substantial effect on the
transition pressures between the crystalline phases of ice at high
pressure \cite{Santra} and one would expect them to matter even more in
solid molecular hydrogen.  An important consequence is that transition
pressures obtained from standard DFT functionals, which neglect vdW
forces, are likely to be unreliable. Because of this recognition, some
very recent work \cite{Morales} has used van der Waals density
functionals to study solid H. There is no need to emphasize that DMC
provides a highly accurate description of exchange and correlation in
relatively weakly correlated materials and that it includes an excellent
quantitative treatment of vdW forces.
 
Combining our static DMC phase diagram and $GW$ results suggests: (a)
that the well known rule of thumb ``the lower the energy the wider the
gap'' \cite{Kaxiras} is valid here; and (b) that metallization occurs
via a transition from the zero band-gap Cmca-12 structure to the
metallic Cmca structure at around 375 GPa. The agreement between the
pressure at which the $GW$ band gap of the Cmca-12 structure closes and
the pressure at which DMC predicts a phase transition from Cmca-12 to
the metallic Cmca structure is remarkable but fortuitous, since
including the effects of phonon zero-point energy changes the picture.

\subsection{Dynamic DMC phase diagram}

The above scenario is based on the static approximation. In reality,
even in the ground state, H$_2$ molecules are quantum oscillators with a
natural frequency of $\sim$$10^{14}$ Hz (for the H--H bond stretching)
and a significant ZPE.  
Following standard practice in the field, our treatment of ZPE uses the
harmonic approximation. A full analysis of anharmonicity is beyond the
scope of this paper, but estimating the ZPE within the harmonic
approximation is feasible \cite{Pickard} and widely believed to produce
accurate results in the case of the molecular solid hydrogen. The
harmonic approximation has also been used for atomic metallic
hydrogen \cite{McMahon}.

To include the effects of ZPE, phonon frequencies were calculated using
density functional perturbation theory as implemented in Quantum
Espresso \cite{QS}. The ZPE per proton was estimated as $E_{ZPE}(V)
= 3\hbar \overline{\omega}/2$, where $\overline{\omega} = \sum_{\bf q}
\sum_{i=1}^{N_{mode}} \omega_{i}({\bf q})/(N_{\bf q}
N_{mode})$.  Here $N_{mode}$ and $N_{\bf q}$ are the
number of vibrational modes in the unit cell and the number of phonon
wave vectors ${\bf q}$, respectively, $\omega_i({\bf q})$ is the angular
frequency of the phonon in band $i$ with wave vector ${\bf q}$, and the
summation over ${\bf q}$ includes all ${\bf k}$-points on a $2 \times 2
\times 2$ grid in the Brillouin zone. McMahon \cite{McMahon}
demonstrated that a ${\bf q}$-point grid of this density is sufficient
to converge ZPE differences between structures to within a few percent.

In agreement with previous DFT results \cite{Pickard}, we find that the
ZPE is large, structure dependent, and density dependent, ranging from
257 meV per proton at low density to 351 meV per proton at high
density. The P$6_3$/m structure has the largest and the Cmca structure
the smallest ZPE over the whole range of pressures considered. The C2/c
and Cmca-12 structures have almost the same ZPE, which lies between the
P$6_3$/m and Cmca values. 

\begin{figure}
\centering
\includegraphics[width=0.50\columnwidth, angle=-90]{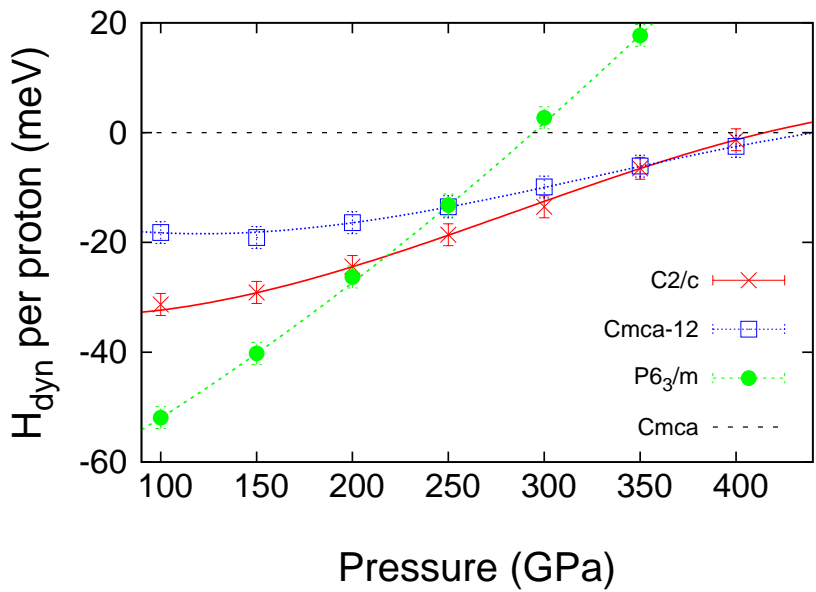} \caption{
  \label{H-dyn-P-stat} (Color online) Enthalpy as a function of
  pressure calculated using the DMC-TABC method in the dynamic
  approximation. The proton ZPE was calculated using DFT. }
\end{figure}

By defining $H_{dyn}=E_{dyn}+PV$, where $E_{dyn} =
E_{stat}+ZPE$, we obtain the dynamic phase diagram of solid
molecular hydrogen illustrated in Fig. ~\ref{H-dyn-P-stat}.

Bearing in mind the error bars shown in Fig. ~\ref{H-dyn-P-stat}, we see that
P$6_3$/m is the most stable phase up to a pressure of approximately 220
GPa. This is in good agreement with experimental results \cite{Akahama}.
As discussed above, because of the tiny enthalpy difference between the
C2/c and Cmca-12 phases, it is difficult to determine their pressure
stability ranges accurately. However, we are able to conclude that we
expect to observe the following structural transitions with increasing
pressure: P$6_3$/m $\rightarrow$ C2/c $\rightarrow$ Cmca-12 $\rightarrow$ Cmca.

Comparing the static and dynamic results, we see that the proton ZPE has
two main effects.  First, although the order in which the phases appear
with increasing pressure is unaffected, their enthalpies and regions of
stability change.  If we were to take an optimistic view and ignore the
error bars, we would say that P$6_3$/m, C2/c, Cmca-12, and Cmca phases
are now stable in the pressure ranges $<$220 GPa, 220--360 GPa, 360--430
GPa, and $>$430 GPa, respectively. As in the static case, however, the
error bars make it difficult to discriminate between the C2/c and
Cmca-12 structures over a wide range of pressures, this time from 300 to
approximately 450 GPa. Second, the inclusion of ZPE increases the
pressure of the transition to the metallic Cmca phase by $14\%$, which
is close to the $10\%$ increase previously reported \cite{Martin}. If
the band gap is of order 0.5 eV (the vibron energy) or less, which is
comparable to the Cmca-12 gap in this pressure range, the inclusion of
ZPE effects is also expected to increase the gap. \cite{Martin} The
increase is small, however, and the gap still closes before the
transition to the Cmca structure takes place.  This leaves a region of
20--30 GPa where the Cmca-12 phase persists but is no longer insulating.
We, also, found that the ZP pressures are quite small and has no significant effect
on the phase diagram. 

A comparison of the densities of different structures at the same
pressure shows that, in phases II and III: density(Cmca) $>$
density(Cmca-12) $>$ density(C2/c) $>$ density(P$6_3$/m).  As the
pressure is increased, the $PV$ term in the enthalpy rises faster for
the low-density insulating phases than for the metallic phase, reducing
the relative stability of the insulating phases and leading eventually
to the metal-insulator transition. The difference between the densities
of the insulating phases and the metallic Cmca phase reduces at high
pressure, especially in the case of the Cmca-12 structure.

The differences between the DMC and DFT phase diagrams are primarily due
to the internal energy term in the enthalpy. For pressures up to 220
GPa, the DMC energy of the low density P$6_3$/m phase lies further below
the energies of the other phases, especially the high-density metallic
Cmca phase, than does the DFT energy. Our results, at both the static
and dynamic levels, predict that the P$6_3$/m phase is the most stable
of the structures considered for a large range of pressures up to 220
GPa in the dynamic case or 280 GPa in the static case. This prediction
is consistent with the latest low temperature X-ray powder-diffraction
results \cite{Akahama}, which indicate that the hydrogen molecules in
phases II and III are still in a hexagonal structure and that phase III
remains hexagonal up to the maximum experimental pressure of 183
GPa. The nature of the difference between phases II and III remains to
be established, but may be related to a classical orientational ordering
of the H$_2$ molecules \cite{Mazin}.

\section{Conclusion}

In conclusion, we have used DMC simulations to study the phase diagram
of the four most stable structures of high pressure hydrogen found from
DFT calculations. We accurately computed the FS corrections by correctly
combining the use of TABC with the DFT-KZK method. We also carried out
$GW$ calculations to find the metallization pressure. Our $GW$ results
show that the band gap of the insulating Cmca-12 phase closes at 373
GPa, whilst our DMC simulations in the dynamic approximation show that
the Cmca-12 structure may remain stable up to 430 GPa, at which point it
transforms to the metallic Cmca phase. We propose that the semimetallic
state observed experimentally at around 360 GPa \cite{Hemley12} may
correspond to the Cmca-12 structure near the pressure at which the band
gap closes. At high pressures our DMC results are in qualitative
agreement with previous DFT calculations \cite{Pickard}, and suggest
that the metallization of solid molecular hydrogen happens through the
transformation of the Cmca-12 phase to the Cmca structure. Our DMC
results indicate that the hcp structure P$6_3$/m, which has the lowest
density and the largest band gap of the phases considered, is the most
stable molecular structure up to 220 GPa.

We acknowledge interesting and helpful discussions with Richard Needs
and Chris Pickard.

\end{document}